\documentclass[12pt, draftclsnofoot, onecolumn]{IEEEtran}
\IEEEoverridecommandlockouts

\newtheorem{proposition}{{Proposition}}

\newtheorem{theorem}{{Theorem}}

\newtheorem{remark}{{Remark}}

\usepackage{cite}
\usepackage{amsmath,amssymb,amsfonts}
\usepackage{algorithmic}
\usepackage{graphicx}
\usepackage{textcomp}
\usepackage{xcolor}
\def\BibTeX{{\rm B\kern-.05em{\sc i\kern-.025em b}\kern-.08em
    T\kern-.1667em\lower.7ex\hbox{E}\kern-.125emX}}
\begin{document}

\title{On the Optimality of the Greedy Policy for Battery Limited Energy Harvesting Communications
}

\author{Ye Wang, Ali Zibaeenejad, Yaohui Jing, Jun Chen 
}

\maketitle

\begin{abstract}
Consider a battery limited energy harvesting communication system with online power control. Assuming independent and identically distributed (i.i.d.) energy arrivals and the harvest-store-use architecture, it is shown that the greedy policy achieves the maximum throughput if and only if the battery capacity is below a certain positive threshold that admits a precise characterization. Simple lower and upper bounds on this threshold are established. The asymptotic relationship between the threshold and the mean of the energy arrival process is analyzed for several examples.



\end{abstract}


\section{Introduction}



The problem of power control for energy harvesting communications has received significant attention in recent years \cite{SMJG10,OTYUY11,YU12,TY12,HZ12,OU12,BGD13,WL13,SK13,XZ14,RSV14,UYESZGH15, DFO15,SO16,ABU18,ZP18,ZP182}. Though the exact problem formulation varies depending on the system model and the performance metric, the essential challenge remains the same, which is, roughly speaking, to deal with random energy availability. In this paper we consider online power control for a battery limited energy harvesting communication system with the goal of maximizing the long-term average throughput. The aforementioned challenge is arguably most pronounced in this setting. Indeed, it is known that the impact of random energy arrivals can be smoothed out if the system is equipped with a battery of unlimited capacity \cite{OU12,ZP18}, and offline power control can achieve the same effect to a certain extent. The standard approach to the problem under consideration is based on the theory of Markov decision processes \cite{ABFGM93}. Although in principle the maximum throughput and the associated optimal online power control policy can be found by solving the relevant Bellman equation,
it is often very difficult to accomplish this task analytically. To the best of our knowledge, there is no exact characterization of the maximum throughput except for Bernoulli energy arrivals \cite{SO16}. To circumvent this difficulty, we tackle the problem from a different angle. Specifically, instead of directly solving the Bellman equation to get the optimal power control policy, we use it to check whether a given power control policy is optimal. This strategy effectively turns a hard optimization problem into a simple decision problem for which more conclusive results can be obtained (see \cite{TCDS17} for the application of a similar strategy in a different context). In particular, it enables us to establish a sufficient and necessary condition for the optimality of the greedy policy, yielding an exact characterization of the maximum throughput in the low-battery-capacity regime.





The rest of the paper is organized as follows. We state  the main results in Section \ref{sec:main} and present the proofs in Section \ref{sec:proof}. Section \ref{sec:example} contains the asymptotic analysis for several illustrative examples. We conclude the paper in Section \ref{sec:conclusion}. 
Throughout this paper, little-$o$ notation 
$f(x)=o_{x\downarrow 0}(\psi(x))$ ($f(x)=o_{x\uparrow\infty}(\psi(x))$) means $\lim_{x\downarrow 0}\frac{f(x)}{\psi(x)}=0$ ($\lim_{x\uparrow\infty}\frac{f(x)}{\psi(x)}=0$),
and the base of the logarithm function is $e$.





\section{Main Results}\label{sec:main}

Consider a discrete-time energy harvesting communication system equipped with a battery of capacity $c$. Let $X(t)$ denote the amount of energy harvested at time $t$, $t=1,2,\cdots$, where $\{X_t\}_{t=1}^\infty$ are assumed to be i.i.d. copies of a nonnegative random variable $X$. An online power control policy is a sequence of mappings $\{f_t\}_{t=1}^{\infty}$ specifying the level of energy consumption $G_t$ in time slot $t$ based on $X^t\triangleq(X_1,\cdots,X_t)$ for all $t$:
\begin{align*}
G_t=f_t(X^t),\quad t=1,2,\cdots.
\end{align*}
Let $B_t$ denote the amount of energy stored in the battery at the beginning of time slot $t$. We have\footnote{Here we adopt the popular harvest-store-use architecture, which should be contrasted with the harvest-use-store architecture in \cite{RSV14}.}
\begin{align*}
B_t=\min\{B_{t-1}-G_{t-1}+X_t,c\},\quad t=1,2,\cdots,
\end{align*}
where $B_{0}\triangleq 0$ and $G_0\triangleq 0$. An online power control policy is said to be admissible if 
\begin{align*}
G_t\leq B_t,\quad t=1,2,\cdots.
\end{align*}
The throughput induced by policy $\{f_t\}_{t=1}^{\infty}$ is defined as
\begin{align*}
\gamma(c)\triangleq\liminf\limits_{n\uparrow\infty}\frac{1}{n}\mathbb{E}\left[\sum\limits_{t=1}^nr(f_t(X^t))\right],
\end{align*} 
where $r:[0,\infty)\rightarrow[0,\infty)$ is a reward function that specifies the instantaneous rate achievable with the given level of energy consumption.
The maximum throughput is defined as
\begin{align*}
\gamma^*(c)\triangleq\sup \gamma(c),
\end{align*}
where the supremum is taken over all admissible online power control policies.

In this paper, we assume that $r$ is a monotonically increasing concave  function
with continuous first-order derivative $r'$. Special attention is paid to the case
\begin{align}
r(x)=\frac{1}{2}\log(1+x),\quad x\geq 0,\label{eq:awgn}
\end{align}
which is relevant to the scenario where the underlying communication system is capacity-achieving for additive Gaussian noise channels.

An online power control policy $\{f_t\}_{t=1}^\infty$ is said to be stationary  if the resulting $\{G_t\}_{t=1}^\infty$ and $\{B_t\}_{t=1}^\infty$ satisfy
$G_t=f(B_t)$, $t=1,2,\cdots$, for some time-invariant function $f$. 
 The following Bellman equation provides an implicit characterization of the maximum throughput and the associated optimal power control policy.
 

\begin{proposition}[Bellman Equation \cite{SO16}]\label{prop:bellman}
	If there exist a nonnegative scalar $\gamma$ and a bounded function $h:[0,c]\rightarrow[0,\infty)$ that satisfy
	\begin{align}
	\gamma+h(b)=\sup\limits_{g\in[0,b]}\{r(g)+\mathbb{E}[h(\min\{b-g+X,c\})]\}\label{eq:Bellman}
	\end{align}
	for all $b\in[0,c]$, then $\gamma^*(c)=\gamma$; moreover, every stationary policy $f$ such that $f(b)$ attains the supremum in (\ref{eq:Bellman}) for all $b\in[0,c]$ is throughput-optimal.
\end{proposition}



\begin{figure*}[htbp]
	\centerline{\includegraphics[width=19cm]{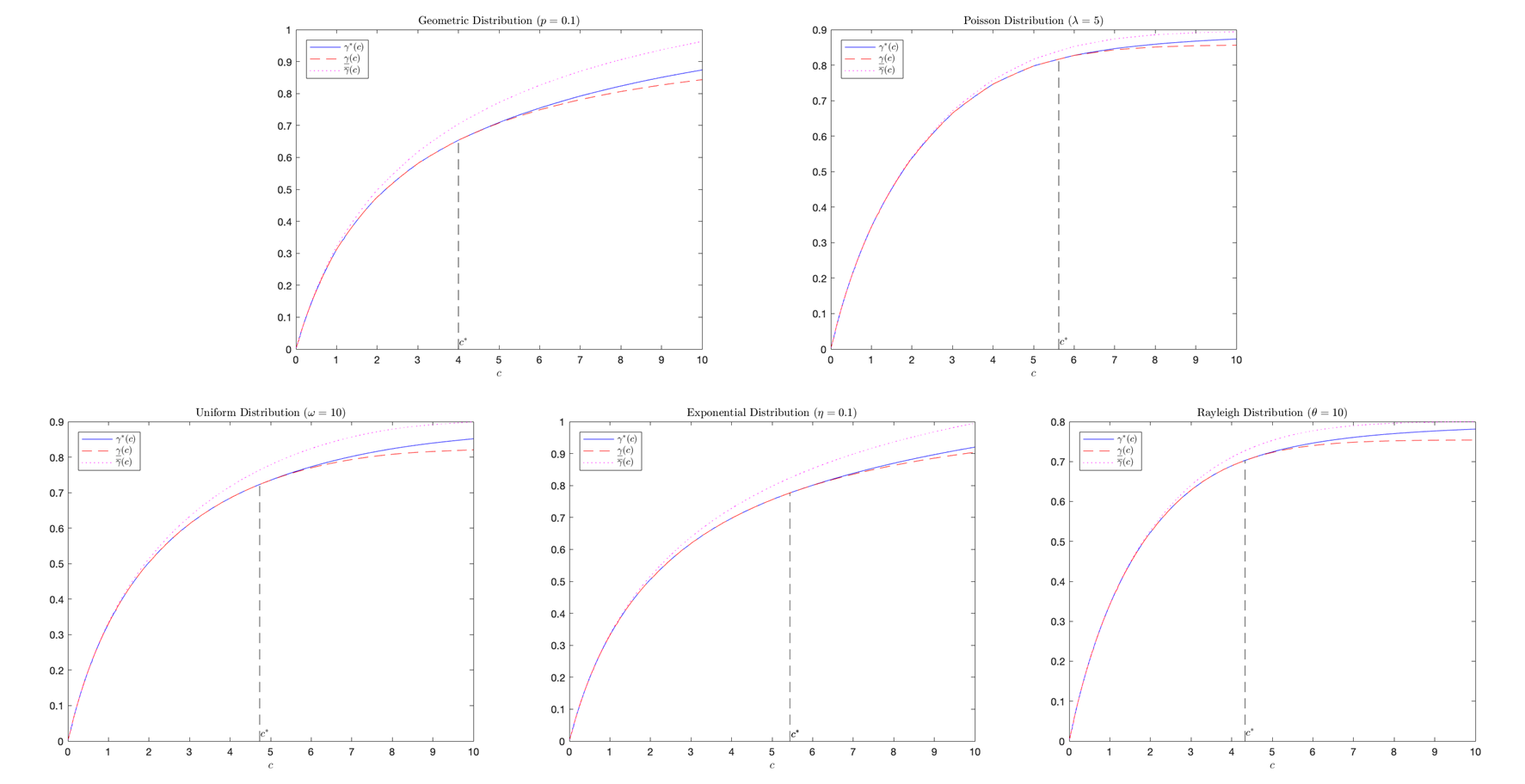}}
	\caption{Illustrations of $\gamma^*(c)$, $\underline{\gamma}(c)$, and $\overline{\gamma}(c)$ for  several different distributions.}
	\label{fig:examples}
\end{figure*}

The greedy policy is a simple stationary policy of the form
\begin{align*}
G_t=B_t,\quad t=1,2,\cdots.
\end{align*}
The throughput induced by the greedy policy can serve as a lower bound on $\gamma^*(c)$:
\begin{align*}
\gamma^*(c)\geq\underline{\gamma}(c)\triangleq\mathbb{E}[r(\min\{X,c\})]. 
\end{align*}
On the other hand, the concavity of the reward function implies the following upper bound on $\gamma^*(c)$ \cite{SO16}:
\begin{align*}
\gamma^*(c)\leq\overline{\gamma}(c)\triangleq r(\mathbb{E}[\min\{X,c\}]).
\end{align*}
Let $\rho(x)\triangleq\mathbb{P}(X<x)$,  $\underline{x}\triangleq\max\{x\geq 0:\rho(x)=0\}$, $\overline{x}\triangleq\inf\{x\geq0:\rho(x)=1\}$, and  $\mu\triangleq\mathbb{E}[X]$. 
We shall assume\footnote{We let $r'(\infty)\triangleq\lim_{x\uparrow\infty}r'(x)$, which is well-defined since $r'$ is a monotonically decreasing function.}  $r'(\underline{x})>r'(\overline{x})$ since otherwise $\underline{\gamma}(c)=\overline{\gamma}(c)$ for all $c\geq 0$. 
It is clear that
\begin{align*}
\lim\limits_{c\downarrow 0}\frac{\underline{\gamma}(c)}{\overline{\gamma}(c)}=1.
\end{align*}
In other words, the greedy policy is asymptotically optimal when $c\downarrow0$. To gain a better understanding, we plot\footnote{Here $\gamma^*(c)$ is obtained by numerically solving the Bellman equation (i.e., (\ref{eq:Bellman})).} $\gamma^*(c)$, $\underline{\gamma}(c)$, and $\overline{\gamma}(c)$ associated with the reward function defined in (\ref{eq:awgn}) 
for various distributions\footnote{The definition of these distributions can be found in Section \ref{sec:example}.} of $X$. It can be seen from the examples in Fig. \ref{fig:examples} that, somewhat surprisingly, $\underline{\gamma}(c)$ coincides with $\gamma^*(c)$ when $c$ is below a certain positive threshold $c^*$ (as a consequence, the greedy policy is in fact exactly optimal in that regime). This turns out to be a general phenomenon, as shown by the following result, which also provides an analytical characterization of $c^*$. 

\begin{theorem}[Threshold $c^*$]\label{thm:main}
	The greedy policy is optimal, i.e., $\gamma^*(c)=\underline{\gamma}(c)$, if and only if $c\leq c^*$, where
	\begin{align*}
	c^*\triangleq\max\{c\geq 0:r'(c)\geq\rho(c)\mathbb{E}[r'(X)|X<c]\}.
	\end{align*}
	In particular, for the reward function defined in (\ref{eq:awgn}),	
	\begin{align}
	c^*=\max\left\{c\geq 0:\frac{1}{1+c}\geq\rho(c)\mathbb{E}\left[\left.\frac{1}{1+X}\right|X<c\right]\right\}.\label{eq:Gthreshold}
	\end{align}
\end{theorem}

\begin{remark}\label{rem:main}
	It is easy to see that $r'(c)$ is a monotonically decreasing continuous function of $c$, and $\rho(c)\mathbb{E}[r'(X)|X<c]$ is a monotonically increasing left-continuous function of $c$; moreover,
	\begin{align*}
	&r'(\underline{x})>\mathbb{P}(X=\underline{x})r'(\underline{x})=\lim\limits_{c\downarrow \underline{x}}\rho(c)\mathbb{E}[r'(X)|X<c],\\
	&r'(\overline{x})<\mathbb{E}[r'(X)]=\lim\limits_{c\downarrow \overline{x}}\rho(c)\mathbb{E}[r'(X)|X<c],\quad \underline{x}<\infty,\\
	&r'(\overline{x})<\mathbb{E}[r'(X)]=\lim\limits_{c\uparrow \overline{x}}\rho(c)\mathbb{E}[r'(X)|X<c],\quad \underline{x}=\infty.
	\end{align*}
	These facts imply that $c^*$ is well-defined and more generally
	\begin{align*}
	\{c\geq 0:r'(c)\geq\rho(c)\mathbb{E}[r'(X)|X<c]\}=[0,c^*]
	\end{align*}
	with $\underline{x}<c^*\leq\overline{x}$ (the second inequality is strict if $\overline{x}=\infty$).
	
	\begin{figure*}[htbp]
		\centerline{\includegraphics[width=18cm]{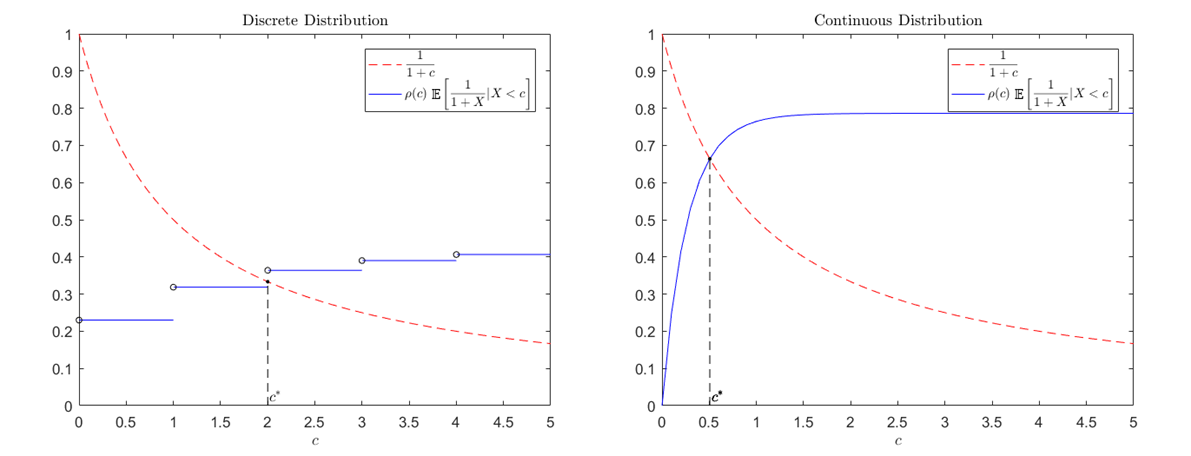}}
		\caption{Characterization of $c^*$ for the case where $X$ has a discrete distribution and the case where $X$ has a continuous distribution.}
		\label{fig:intersection}
	\end{figure*}
	
	\begin{remark}
	To gain a deeper understanding of (\ref{eq:Gthreshold}), it is instructive to consider the following two cases separately (see also Fig. \ref{fig:intersection}).

 1) Let $X$ be a discrete random variable with probability mass function $p_X$. For simplicity, we assume the support of $p_X$ is a countable set $\{\xi_1,\xi_2,\cdots\}$ with $0\leq \xi_1<\xi_2<\cdots$. In this case, $c^*$ is the unique positive number satisfying one of the following two conditions.
	\begin{enumerate}
		\item[i)] $c^*\in(\xi_j,\xi_{j+1})$ for some $j$ and
		\begin{align*}
		\frac{1}{1+c^*}=\sum\limits_{i=1}^j\frac{1}{1+\xi_i}p_X(\xi_i).
		\end{align*}
		\item[ii)] $c^*=\xi_{j+1}$ for some $j$ and
		\begin{align*}
		&\sum\limits_{i=1}^j\frac{1}{1+\xi_i}p_X(\xi_i)\leq\frac{1}{1+c^*}\leq\sum\limits_{i=1}^{j+1}\frac{1}{1+\xi_i}p_X(\xi_i).
		\end{align*}
	\end{enumerate}

2) Let $X$ be a continuous random variable with probability density function $f_X$. In this case, $c^*$ is the unique positive number satisfying
\begin{align}
\frac{1}{1+c^*}=\int_0^{c^*}\frac{1}{1+x}f_X(x)\mathrm{d}x.\label{eq:contin}
\end{align}
\end{remark}

\end{remark}
\begin{IEEEproof}
	See Section \ref{sec:prooftheorem}. Note that for the reward function defined in (\ref{eq:awgn}), 	
	\begin{align*}
	r'(x)=\frac{1}{2(1+x)},\quad x\geq 0,
	\end{align*}
	from which (\ref{eq:Gthreshold}) follows immediately. 
\end{IEEEproof}


Next we establish bounds on $c^*$ that are in general easier to evaluate than $c^*$ itself. For $c>\underline{x}$, let $\overline{r'}_{[\underline{x},c]}$ ($\underline{r'}_{[\underline{x},c]}$) denote the upper concave envelope (the lower convex envelope) of $r'$ over $[\underline{x},c]$.

\begin{proposition}[Lower Bound on $c^*$]\label{cor:lower}
	\begin{align}
c^*\geq\underline{c}\triangleq\sup\{c\in(\underline{x},\overline{x}): r'(c)\geq\rho(c)\overline{r'}_{[\underline{x},c]}(\underline{\xi})
\},\label{eq:underlinec}
\end{align}
where
\begin{align*}
\underline{\xi}\triangleq\max\left\{\frac{\mu-(1-\rho(c))\overline{x}}{\rho(c)},\underline{x}\right\}.
\end{align*}
In particular, for the reward function defined in (\ref{eq:awgn}),
\begin{align}
\underline{c}=\sup\{c\in(\underline{x},\overline{x}):c\leq\overline{\zeta}(c)\},\label{eq:Gunderlinec}
\end{align}
where
\begin{align*}
\overline{\zeta}(c)\triangleq\frac{(1-\rho(c))(1+\underline{x})+\rho(c)\underline{\xi}}{\rho(c)}.
\end{align*}
\end{proposition}
\begin{remark}\label{rem:lowerbound}
	It is clear that 
	\begin{align*}
	&r'(\underline{x})>\mathbb{P}(X=\underline{x})r'(\underline{x})=\lim\limits_{c\downarrow\underline{x}}\rho(c)\overline{r'}_{[\underline{x},c]}(\underline{\xi}),\\
	&r'(\overline{x})<r'(\underline{x})=\lim\limits_{c\uparrow\overline{x}}\rho(c)\overline{r'}_{[\underline{x},c]}(\underline{\xi}),\quad\overline{x}=\infty.
	\end{align*}
	Therefore, we must have $\underline{x}<\underline{c}\leq\overline{x}$ (the second inequality is strict if $\overline{x}=\infty$).
	
	
\end{remark} 
\begin{IEEEproof}
	See Section \ref{sec:proofproposition1}. Note that for the reward function defined in (\ref{eq:awgn}),
	\begin{align*}
	\overline{r'}_{[\underline{x},c]}(x)=\frac{1+\underline{x}+c-x}{2(1+\underline{x})(1+c)},\quad x\in[\underline{x},c],
	\end{align*}
	from which (\ref{eq:Gunderlinec}) follows immediately.
\end{IEEEproof}

\begin{proposition}[Upper Bound on $c^*$]\label{cor:upper}
		\begin{align}
	c^*\leq\overline{c}\triangleq\sup\{c\in(\underline{x},\overline{x}): r'(c)\geq\rho(c)\underline{r'}_{[\underline{x},c]}(\overline{\xi})
	\},\label{eq:overlinec}
	\end{align}
	where
	\begin{align*}
	\overline{\xi}\triangleq\min\left\{\frac{\mu-(1-\rho(c))c}{\rho(c)},c\right\}.
	\end{align*}
	In particular, for the reward function defined in (\ref{eq:awgn}),
	\begin{align}
	\overline{c}=\sup\left\{c\in(\underline{x},\overline{x}): c\leq\underline{\zeta}(c)\right\},\label{eq:Goverlinec}
	\end{align}
	where
	\begin{align*}
	\underline{\zeta}(c)\triangleq\frac{1-\rho(c)+\overline{\xi}}{\rho(c)}.
	\end{align*}
		\end{proposition}
	\begin{remark}\label{remark:overlinec}
		It is clear that 
		\begin{align*}
		&r'(\mu)>\mathbb{P}(X\leq\mu)r'(\mu)=\lim\limits_{c\downarrow\mu}\rho(c)\underline{r'}_{[\underline{x},c]}(\overline{\xi}).
		\end{align*}
		Therefore, we must have $\mu<\overline{c}\leq\overline{x}$. This implies that ``$c\in(\underline{x},\overline{x})$" in (\ref{eq:overlinec}) and (\ref{eq:Goverlinec})  can be replaced by ``$c\in(\mu,\overline{x})$".		
		In particular, we can write (\ref{eq:Goverlinec}) equivalently as
		\begin{align*}
		\overline{c}=\sup\left\{c\in(\mu,\overline{x}): c\leq\frac{\mu+\rho(c)-\rho^2(c)}{1-\rho(c)+\rho^2(c)}\right\}.
		\end{align*}
		Note that $\overline{c}=\overline{x}$ may hold even if $\overline{x}=\infty$. As shown in Appendix \ref{app:overlinec},  $\overline{c}=\infty$ if $\overline{x}=\infty$ and $r'(\mu-\epsilon)=r'(\overline{x})$ for some $\epsilon>0$; on the other hand, if $\overline{x}=\infty$ and $r'(\mu)>r'(\overline{x})$, then $\overline{c}<\infty$.


				

		
		
		

	\end{remark}
\begin{IEEEproof}
	See Section \ref{sec:proofupper}. Note that for the reward function defined in (\ref{eq:awgn}),
	\begin{align*}
	\underline{r'}_{[\underline{x},c]}(x)=\frac{1}{2(1+x)},\quad x\in[\underline{x},c],
	\end{align*}
	from which (\ref{eq:Goverlinec}) follows immediately.
\end{IEEEproof}

We further establish semi-universal bounds on $c^*$ that depend only on $\underline{x}$, $\overline{x}$, and $\mu$.

\begin{proposition}[Semi-Universal Lower Bound on $c^*$]\label{cor:lower2}
\begin{align}
c^*\geq\underline{\underline{c}}\triangleq\sup\{c\in(\underline{x},\overline{x}):r'(c)\geq\overline{\chi}(c)
\},\label{eq:doubleunderlinec}
\end{align}
where
\begin{align*}
\overline{\chi}(c)\triangleq
\begin{cases}
\sup\limits_{\rho(c)\in\left(0,\frac{\overline{x}-\mu}{\overline{x}-c}\right)}\rho(c)\overline{r'}_{[\underline{x},c]}(\underline{\xi}),& c\in(\underline{x},\mu],\\
\sup\limits_{\rho(c)\in\left(\frac{c-\mu}{c-\underline{x}},1\right)}\rho(c)\overline{r'}_{[\underline{x},c]}(\underline{\xi}), & c\in(\mu,\overline{x}).
\end{cases}
\end{align*}
	In particular, for the reward function defined in (\ref{eq:awgn}),
	\begin{align}
	\underline{\underline{c}}=\begin{cases}
	\frac{(1+\underline{x})(\overline{x}-\underline{x})}{\overline{x}-\mu}-1, &\mu\leq\overline{x}-\underline{x}-1,\\
	\mu, &\mu>\overline{x}-\underline{x}-1.
	\end{cases} \label{eq:Gdoubleunderlinec}
	\end{align}
\end{proposition}
\begin{remark}
	We let $\underline{\underline{c}}\triangleq\underline{x}$ if $\{c\in(\underline{x},\overline{x}):r'(c)\geq\overline{\chi}(c)
		\}=\emptyset$.
\end{remark}
\begin{IEEEproof}
	See Section \ref{sec:proofproposition3}.
\end{IEEEproof}

\begin{proposition}[Semi-Universal Upper Bound on $c^*$]\label{cor:upper2}
	\begin{align}
	c^*\leq\overline{\overline{c}}\triangleq\sup\{c\in(\underline{x},\overline{x}):r'(c)\geq\underline{\chi}(c)
	\},\label{eq:doubleoverlinec}
	\end{align}
	where
	\begin{align*}
	\underline{\chi}(c)\triangleq
	\begin{cases}
	\inf\limits_{\rho(c)\in\left(0,\frac{\overline{x}-\mu}{\overline{x}-c}\right)}\rho(c)\underline{r'}_{[\underline{x},c]}(\overline{\xi}),& c\in(\underline{x},\mu],\\
	\inf\limits_{\rho(c)\in\left(\frac{c-\mu}{c-\underline{x}},1\right)}\rho(c)\underline{r'}_{[\underline{x},c]}(\overline{\xi}), & c\in(\mu,\overline{x}).
	\end{cases}
	\end{align*}
	In particular, for the reward function defined in (\ref{eq:awgn}),
\begin{align}
\overline{\overline{c}}=
\begin{cases}
\min\{c_1,\overline{x}\},&\mu\leq\frac{3}{2}\underline{x}+\frac{1}{2},\\
\min\{c_2,\overline{x}\},&\mu>\frac{3}{2}\underline{x}+\frac{1}{2},
\end{cases}
\label{eq:Gdoubleoverlinec}
\end{align}
where
\begin{align*}
&c_1\triangleq\frac{\mu+\underline{x}+\sqrt{(\mu+\underline{x})^2-4(\underline{x}^2+\underline{x}-\mu)}}{2},\\
&c_2\triangleq\frac{4}{3}\mu+\frac{1}{3}.
\end{align*}
\end{proposition}
\begin{IEEEproof}
See Section \ref{sec:proofproposition4}.
\end{IEEEproof}

\begin{figure}[htbp]
	\centerline{\includegraphics[width=9cm]{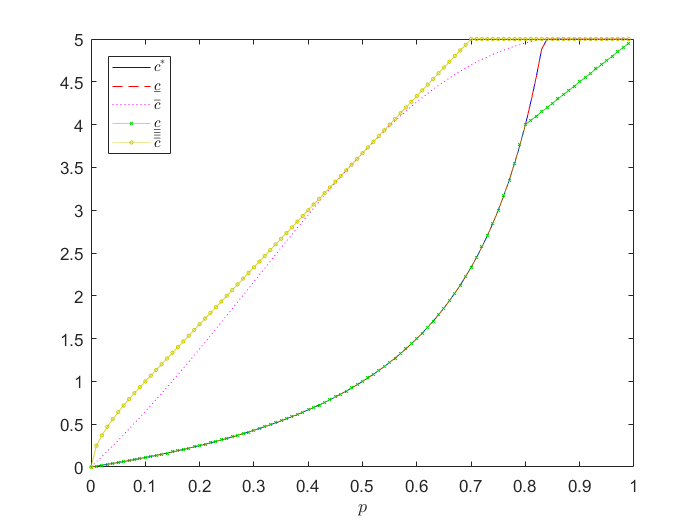}}
	\caption{Plots of $c^*$, $\underline{c}$, $\overline{c}$, $\underline{\underline{c}}$, and $\overline{\overline{c}}$ against $p$ with $\underline{x}=0$ and $\overline{x}=5$.}
	\label{fig:bound}
\end{figure}

Consider the reward function defined in (\ref{eq:awgn}) and assume that $X$ is a Bernoulli random variable with $\mathbb{P}(X=\underline{x})=1-p$ and $\mathbb{P}(X=\overline{x})=p$, where  $p\in(0,1)$. For this special example, a simple calculation shows that
\begin{align*}
&c^*=\underline{c}=\begin{cases}
\frac{\underline{x}+p}{1-p},&  \frac{\underline{x}+p}{1-p}\leq\overline{x},\\
\overline{x}, &  \frac{\underline{x}+p}{1-p}>\overline{x},
\end{cases}\\
&\overline{c}=\begin{cases}
\frac{(1-p)(\underline{x}+p)+p\overline{x}}{1-p+p^2},& \frac{\underline{x}+p}{1-p}\leq\overline{x},\\
\overline{x},&\frac{\underline{x}+p}{1-p}>\overline{x},
\end{cases}\\
&\underline{\underline{c}}=\begin{cases}
\frac{\underline{x}+p}{1-p}, &\frac{(2-p)\underline{x}+1}{1-p}\leq\overline{x},\\
(1-p)\underline{x}+p\overline{x}, &\frac{(2-p)\underline{x}+1}{1-p}>\overline{x},
\end{cases}\\
&\overline{\overline{c}}=\begin{cases}
\min\{c_1,\overline{x}\},&\frac{(1+2p)\underline{x}+1}{2p}\geq\overline{x},\\
\min\{c_2,\overline{x}\},&\frac{(1+2p)\underline{x}+1}{2p}<\overline{x},
\end{cases}
\end{align*}
where
\begin{align*}
c_1&=\frac{(2-p)\underline{x}+p\overline{x}}{2}+\frac{\sqrt{((2-p)\underline{x}+p\overline{x})^2-4(\underline{x}^2+p(\underline{x}-\overline{x}))}}{2},\\
c_2&=\frac{4}{3}((1-p)\underline{x}+p\overline{x})+\frac{1}{3}.
\end{align*}
Moreover, it can be verified that
$\overline{c}=c^*$ if  $\frac{\underline{x}+p}{1-p}\geq\overline{x}$, $\underline{\underline{c}}=c^*$ if $\frac{(2-p)\underline{x}+1}{1-p}\leq\overline{x}$, and $\overline{\overline{c}}=c^*$ if $\overline{x}\leq\min\{\frac{(1+2p)\underline{x}+1}{2p},c_1\}$ or $\frac{(3-4p)\overline{x}-1}{4(1-p)}\leq\underline{x}\leq\frac{2p\overline{x}-1}{1+2p}$. Therefore, the bounds in Propositions \ref{cor:lower}, \ref{cor:upper}, \ref{cor:lower2}, and \ref{cor:upper2} are tight for non-trivial cases. Plots of $c^*$, $\underline{c}$, $\overline{c}$, $\underline{\underline{c}}$, and $\overline{\overline{c}}$ against $p$ with $\underline{x}=0$ and $\overline{x}=5$ can be found in Fig. \ref{fig:bound}.

\section{Proofs}\label{sec:proof}

\subsection{Proof of Theorem \ref{thm:main}}\label{sec:prooftheorem}

The main difficulty in solving the Bellman equation (i.e., (\ref{eq:Bellman})) is that the function $h$ associated with the optimal power control policy is in general unknown. 
However, since we only aim to check the optimality of the greedy policy, it is easy to construct a candidate function $h$. Specifically,  in view of Proposition \ref{prop:bellman}, the greedy policy is optimal if
\begin{align*}
\sup\limits_{g\in[0,b]}\{r(g)+\mathbb{E}[h(\min\{b-g+X,c\})]\}=r(b)+\mathbb{E}[h(\min\{X,c\})]=\underline{\gamma}(c)+h(b)
\end{align*}
for all $b\in[0,c]$, and the second equality naturally suggests that
$h(x)=r(x)$ for $x\in[0,c]$.
Therefore, it suffices to check whether the supremum of $\phi(g)\triangleq r(g)+\mathbb{E}[r(\min\{b-g+X,c\})]$ over $[0,b]$ is attained at $g=b$ for all $b\in[0,c]$. We show in Appendix \ref{app:leftright} that for $g\in(0,b]$, 
\begin{align}
\lim\limits_{\epsilon\downarrow 0}\frac{1}{\epsilon}(\phi(g)-\phi(g-\epsilon))=r'(g)-\rho(c-b+g)\mathbb{E}[r'(b-g+X)|X<c-b+g],\label{eq:left}
\end{align}
and for $g\in[0,b)$,
\begin{align}
&\lim\limits_{\epsilon\downarrow 0}\frac{1}{\epsilon}(\phi(g+\epsilon)-\phi(g))\nonumber\\
&=r'(g)-\rho(c-b+g)\mathbb{E}[r'(b-g+X)|X<c-b+g]-\mathbb{P}(X=c-b-g)r'(c).\label{eq:right}
\end{align}
Therefore, $\phi$ is semi-differentiable and consequently is continuous over $[0,b]$. Note that for $c\in[0,c^*]$, $b\in[0,c]$, and $g\in[0,b]$,
\begin{align*}
r'(g)-\rho(c-b+g)\mathbb{E}[r'(b-g+X)|X<c-b+g]\geq r'(c)-\rho(c)\mathbb{E}[r'(X)|X<c]\geq 0.
\end{align*}
So $\phi$ is a monotonically increasing function\footnote{Here we have invoked the fact that a continuous function $f$ with nonnegative left derivative must be monotonically increasing. This fact can be proved as follows. Assume there exist $\alpha<\beta$ such that $f(\alpha)>f(\beta)$. Let $\kappa\triangleq\frac{f(\alpha)-f(\beta)}{2(\beta-\alpha)}$ and  $\tau\triangleq\max\{x\in[\alpha,\beta]:f(x)-f(\beta)>\kappa(\beta-x)\mbox{ for all }x'\in[\alpha,x)\}$. It follows by the continuity of $f$ that $\tau\in(\alpha,\beta]$ and $f(\tau)-f(\beta)=\kappa(\beta-\tau)$. Since the left derivative of $f$ is nonnegative at $\tau$, there exists $\tau'\in[\alpha,\tau)$ such that $f(\tau')-f(\tau)\leq\kappa(\tau-\tau')$. Therefore, we have $f(\tau')-f(\beta)=f(\tau')-f(\tau)+f(\tau)-f(\beta)\leq\kappa(\beta-\tau')$, which is contradictory to the definition of $\tau$.
} over $[0,b]$ for all $b\in[0,c]$ when $c\leq c^*$. This proves the ``if" part of Theorem \ref{thm:main}.





To prove the ``only if" part of Theorem \ref{thm:main}, we shall construct an online power control policy that outperforms the greedy policy when $c>c^*$. To this end, we modify the greedy policy as follows: for $t=1,2,\cdots$,
\begin{align*}
&G_{2t-1}=
\begin{cases}
B_{2t-1}-\epsilon, & X_{2t-1}\geq \min\{\overline{x},c\}-\epsilon,\\
B_{2t-1}, & \mbox{otherwise},
\end{cases}\\
&G_{2t}=B_{2t},
\end{align*}
where $\epsilon\in(0,\frac{1}{2}\min\{\overline{x},c\}]$.  As compared to the greedy policy, the modified policy incurs a rate loss $\mathbb{E}[r(\min\{X_{2t-1},c\})]-\mathbb{E}[r(G_{2t-1})]$ in time slot $2t-1$, but gains $\mathbb{E}[r(G_{2t})]-\mathbb{E}[r(\min\{X_{2t},c\})]$ in time slot $2t$. It can be verified that
\begin{align*}
&\mathbb{E}[r(\min\{X_{2t-1},c\})]-\mathbb{E}[r(G_{2t-1})]\\
&=\mathbb{P}(X_{2t-1}\geq\min\{\overline{x},c\}-\epsilon)\mathbb{E}[r(\min\{X_{2t-1},c\})-r(\min\{X_{2t-1},c\}-\epsilon)|X_{2t-1}\geq\min\{\overline{x},c\}-\epsilon]\\
&=\mathbb{P}(\min\{\overline{x},c\}-\epsilon\leq X_{2t-1}<c)\mathbb{E}[r(X_{2t-1})-r(X_{2t-1}-\epsilon)|\min\{\overline{x},c\}-\epsilon\leq X_{2t-1}<c]\\
&\quad+\mathbb{P}(X_{2t-1}\geq c)\mathbb{E}[r(c)-r(c-\epsilon)|X_{2t-1}\geq c]\\
&\leq\mathbb{P}(\min\{\overline{x},c\}-\epsilon\leq X_{2t-1}<c)\mathbb{E}[r'(\min\{\overline{x},c\}-2\epsilon)\epsilon|\min\{\overline{x},c\}-\epsilon\leq X_{2t-1}<c]\\
&\quad+\mathbb{P}(X_{2t-1}\geq c)\mathbb{E}[r'(\min\{\overline{x},c\}-2\epsilon)\epsilon|X_{2t-1}\geq c]\\
&=\mathbb{P}(X\geq\min\{\overline{x},c\}-\epsilon)r'(\min\{\overline{x},c\}-2\epsilon)\epsilon,
\end{align*}
and
\begin{align*}
&\mathbb{E}[r(G_{2t})]-\mathbb{E}[r(\min\{X_{2t},c\})]\\
&=\mathbb{P}(X_{2t-1}\geq\min\{\overline{x},c\}-\epsilon)\mathbb{E}[r(\min\{X_{2t}+\epsilon,c\})-r(\min\{X_{2t},c\})]\\
&=\mathbb{P}(X_{2t-1}\geq\min\{\overline{x},c\}-\epsilon)(\mathbb{P}(X_{2t}< c-\epsilon)\mathbb{E}[r(X_{2t}+\epsilon)-r(X_{2t})|X_{2t}< c-\epsilon]\\
&\qquad+\mathbb{P}(c-\epsilon\leq X_{2t}<c)\mathbb{E}[r(c)-r(X_{2t})|c-\epsilon\leq X_{2t}<c])\\
&\geq\mathbb{P}(X_{2t-1}\geq\min\{\overline{x},c\}-\epsilon)\mathbb{P}(X_{2t}< c-\epsilon)\mathbb{E}[r(X_{2t}+\epsilon)-r(X_{2t})|X_{2t}< c-\epsilon]\\
&\geq\mathbb{P}(X_{2t-1}\geq\min\{\overline{x},c\}-\epsilon)\mathbb{P}(X_{2t}< c-\epsilon)\mathbb{E}[r'(X_{2t}+\epsilon)\epsilon|X_{2t}< c-\epsilon]\\
&=\mathbb{P}(X\geq\min\{\overline{x},c\}-\epsilon)\rho(c-\epsilon)\mathbb{E}[r'(X+\epsilon)|X< c-\epsilon]\epsilon.
\end{align*}
Clearly, we have
\begin{align*}
\mathbb{P}(X\geq\min\{\overline{x},c\}-\epsilon)>0,\quad \epsilon>0.
\end{align*}
Moreover, 
\begin{align*}
&\lim\limits_{\epsilon\downarrow 0}r'(\min\{\overline{x},c\}-2\epsilon)=r'(\min\{\overline{x},c\}),
\end{align*}
and it follows by the monotone convergence theorem that
\begin{align*}
\lim\limits_{\epsilon\downarrow 0}\rho(c-\epsilon)
\mathbb{E}[r'(X+\epsilon)|X< c-\epsilon]=\rho(c)\mathbb{E}[r'(X)|X< c].
\end{align*}
If $c\leq\overline{x}$, 
\begin{align}
r'(\min\{\overline{x},c\})&=r'(c)\nonumber\\
&<\rho(c)\mathbb{E}[r'(X)|X< c],\label{eq:assumption1}
\end{align}
where (\ref{eq:assumption1}) is due to the assumption that $c>c^*$.
If $c>\overline{x}$,
\begin{align}
r'(\min\{\overline{x},c\})&=r'(\overline{x})\nonumber\\
&<\mathbb{E}[r'(X)]\label{eq:assumption2}\\
&=\rho(c)\mathbb{E}[r'(X)|X< c],\nonumber
\end{align}
where (\ref{eq:assumption2}) is due to the assumption that $r'(\underline{x})>r'(\overline{x})$. Therefore, when $\epsilon$ is sufficiently close to 0,
\begin{align*}
\mathbb{E}[r(\min\{X_{2t-1},c\})]-\mathbb{E}[r(G_{2t-1})]<\mathbb{E}[r(G_{2t})]-\mathbb{E}[r(\min\{X_{2t},c\})]
\end{align*}
and the overall throughput is improved. This proves the ``only if" part of Theorem \ref{thm:main}.


\begin{remark}
The proof of the ``if" part can be slightly modified to show that as long as $r'$ is continuous and positive (or constantly zero) over $[0,\nu]$ for some $\nu>0$, the greedy policy is optimal when $c$ is sufficiently close to $0$. Characterizing the sufficient and necessary condition for the optimality of the greedy policy under relaxed assumptions on the reward function is left for future work.	
	
\end{remark}

\begin{remark}
	Intuitively, it makes sense to save energy only when the expected future return exceeds the current loss;
	 with a small battery, one has  no impetus to keep some energy for later because there is a good chance that the next energy arrival by itself will get the battery fully charged, rendering the saved energy wasted. This intuitive explanation also suggests that the optimality of the greedy policy is specific to online power control. Indeed, for offline power control or, more generally, power control with the knowledge of future energy arrivals in a
	  look-ahead window \cite{ZC19}, 	  
	  one can effectively avoid the situation that the saved energy gets wasted due to battery overflow
	  	  and consequently the greedy policy is in general strictly suboptimal (even in the low-battery-capacity regime).
	

\end{remark}

\subsection{Proof of Proposition \ref{cor:lower}}\label{sec:proofproposition1}

	For $c\in(\underline{x},\overline{x})$,
\begin{align}
\mathbb{E}[r'(X)|X<c]&\leq \mathbb{E}[\overline{r'}_{[\underline{x},c]}(X)|X<c]\nonumber\\
&\leq\overline{r'}_{[\underline{x},c]}(\mathbb{E}[X|X<c]),\label{eq:Jesen1}
\end{align}
where (\ref{eq:Jesen1}) is due to Jensen's inequality. Note that
\begin{align*}
\mu=\rho(c)\mathbb{E}[X|X<c]+(1-\rho(c))\mathbb{E}[X|X\geq c]\leq\rho(c)\mathbb{E}[X|X<c]+(1-\rho(c))\overline{x},
\end{align*}
which implies 
\begin{align*}
\mathbb{E}[X|X<c]\geq\frac{\mu-(1-\rho(c))\overline{x}}{\rho(c)}.
\end{align*}
Moreover, we have $\mathbb{E}[X|X<c]\geq\underline{x}$.
Since $\overline{r'}_{[\underline{x},c]}(x)$ is a monotonically decreasing function of $x$ over $[\underline{x},c]$, it follows that
\begin{align}
\overline{r'}_{[\underline{x},c]}(\mathbb{E}[X|X<c])\leq\overline{r'}_{[\underline{x},c]}(\underline{\xi}).\label{eq:monotone1}
\end{align}
Combining (\ref{eq:Jesen1}) and (\ref{eq:monotone1}) gives
\begin{align*}
\mathbb{E}[r'(X)|X<c]\leq\overline{r'}_{[\underline{x},c]}(\underline{\xi}).
\end{align*}
Therefore,
\begin{align*}
\left\{c\in(\underline{x},\overline{x}): r'(c)\geq\rho(c)\overline{r'}_{[\underline{x},c]}(\underline{\xi})
\right\}\subseteq\left\{c\in(\underline{x},\overline{x}): r'(c)\geq\rho(c)\mathbb{E}[r'(X)|X<c]
\right\},
\end{align*}
which, together with the fact (see Remark \ref{rem:main}) that
\begin{align*}
\sup\left\{c\in(\underline{x},\overline{x}): r'(c)\geq\rho(c)\mathbb{E}[r'(X)|X<c]
\right\}=c^*,
\end{align*}
proves (\ref{eq:underlinec}).

\subsection{Proof of Proposition \ref{cor:upper}}\label{sec:proofupper}

	For $c\in(\underline{x},\overline{x})$,
\begin{align}
\mathbb{E}[r'(X)|X<c]&\geq \mathbb{E}[\underline{r'}_{[\underline{x},c]}(X)|X<c]\nonumber\\
&\geq\underline{r'}_{[\underline{x},c]}(\mathbb{E}[X|X<c]),\label{eq:Jesen2}
\end{align}
where (\ref{eq:Jesen2}) is due to Jensen's inequality. Note that
\begin{align*}
\mu=\rho(c)\mathbb{E}[X|X<c]+(1-\rho(c))\mathbb{E}[X|X\geq c]\geq\rho(c)\mathbb{E}[X|X<c]+(1-\rho(c))c,
\end{align*}
which implies 
\begin{align*}
\mathbb{E}[X|X<c]\leq\frac{\mu-(1-\rho(c))c}{\rho(c)}.
\end{align*}
Moreover, we have $\mathbb{E}[X|X<c]\leq c$.
Since $\underline{r'}_{[\underline{x},c]}(x)$ is a monotonically decreasing function of $x$ over $[\underline{x},c]$, it follows that
\begin{align}
\underline{r'}_{[\underline{x},c]}(\mathbb{E}[X|X<c])\geq\underline{r'}_{[\underline{x},c]}(\overline{\xi}).\label{eq:monotone2}
\end{align}
Combining (\ref{eq:Jesen2}) and (\ref{eq:monotone2}) gives
\begin{align*}
\mathbb{E}[r'(X)|X<c]\geq\underline{r'}_{[\underline{x},c]}(\overline{\xi}).
\end{align*}
Therefore,
\begin{align*}
\left\{c\in(\underline{x},\overline{x}): r'(c)\geq\rho(c)\overline{r'}_{[\underline{x},c]}(\underline{\xi})
\right\}\supseteq\left\{c\in(\underline{x},\overline{x}): r'(c)\geq\rho(c)\mathbb{E}[r'(X)|X<c]
\right\},
\end{align*}
which, together with the fact (see Remark \ref{rem:main}) that
\begin{align*}
\sup\left\{c\in(\underline{x},\overline{x}): r'(c)\geq\rho(c)\mathbb{E}[r'(X)|X<c]
\right\}=c^*,
\end{align*}
proves (\ref{eq:overlinec}).

\subsection{Proof of Proposition \ref{cor:lower2}}\label{sec:proofproposition3}

For $c\in(\underline{x},\overline{x})$, we have $\rho(c)\in(0,1)$ and
\begin{align*}
\rho(c)\underline{x}+(1-\rho(c))c<\mu<\rho(c)c+(1-\rho(c))\overline{x},
\end{align*}
which can be written equivalently as
\begin{align*}
\frac{c-\mu}{c-\underline{x}}<\rho(c)<\frac{\overline{x}-\mu}{\overline{x}-c}.
\end{align*}
Therefore, we have 
\begin{align*}
\{c\in(\underline{x},\overline{x}):r'(c)\geq\overline{\chi}(c)
\}\subseteq\{c\in(\underline{x},\overline{x}):r'(c)\geq\rho(c)\overline{r'}_{[\underline{x},c]}(\underline{\xi})
\}
\end{align*}
and consequently $\underline{\underline{c}}\leq\underline{c}$. Invoking Proposition \ref{cor:lower} proves (\ref{eq:doubleunderlinec}).

Now we proceed to prove (\ref{eq:Gdoubleunderlinec}). It suffices to consider the case $\overline{x}<\infty$ since otherwise $\underline{\underline{c}}=\underline{x}$ and (\ref{eq:Gdoubleunderlinec}) is obviously true.	
Clearly, $r'(c)\geq\overline{\chi}(c)$ if and only if
\begin{align*}
c\leq\begin{cases}
\inf\limits_{\rho(c)\in\left(0,\frac{\overline{x}-\mu}{\overline{x}-c}\right)}\overline{\zeta}(c), & c\in(\underline{x},\mu],\\
\inf\limits_{\rho(c)\in\left(\frac{c-\mu}{c-\underline{x}},1\right)}\overline{\zeta}(c), & c\in(\mu,\overline{x}),
\end{cases}
\end{align*}
where
\begin{align*}
\overline{\zeta}(c)=\begin{cases}
\frac{1+\underline{x}}{\rho(c)}-1,&\rho(c)\in\left(0,\frac{\overline{x}-\mu}{\overline{x}-\underline{x}}\right],\\
\frac{\mu-\overline{x}+\underline{x}+1}{\rho(c)}+\overline{x}-\underline{x}-1,&\rho(c)\in\left(\frac{\overline{x}-\mu}{\overline{x}-\underline{x}},1\right).
\end{cases}
\end{align*}
For $c\in(\underline{x},\mu]$,
\begin{align}
\inf\limits_{\rho(c)\in\left(0,\frac{\overline{x}-\mu}{\overline{x}-c}\right)}\overline{\zeta}(c)
&=\min\left\{\inf\limits_{\rho(c)\in\left(0,\frac{\overline{x}-\mu}{\overline{x}-\underline{x}}\right]}\frac{1+\underline{x}}{\rho(c)}-1,\inf\limits_{\rho(c)\in\left(\frac{\overline{x}-\mu}{\overline{x}-\underline{x}},\frac{\overline{x}-\mu}{\overline{x}-c}\right)}\frac{\mu-\overline{x}+\underline{x}+1}{\rho(c)}+\overline{x}-\underline{x}-1\right\}\nonumber\\
&=\inf\limits_{\rho(c)\in\left(\frac{\overline{x}-\mu}{\overline{x}-\underline{x}},\frac{\overline{x}-\mu}{\overline{x}-c}\right)}\frac{\mu-\overline{x}+\underline{x}+1}{\rho(c)}+\overline{x}-\underline{x}-1\label{eq:twoinf}\\
&=\begin{cases}
\frac{(1+\underline{x})(\overline{x}-\underline{x})}{\overline{x}-\mu}-1, &\mu\leq\overline{x}-\underline{x}-1,\\\frac{(\mu-\overline{x}+\underline{x}+1)(\overline{x}-c)}{\overline{x}-\mu}+\overline{x}-\underline{x}-1, &\mu>\overline{x}-\underline{x}-1,
\end{cases}\label{eq:case1a}
\end{align}
where (\ref{eq:twoinf}) is due to the fact that
\begin{align}
\inf\limits_{\rho(c)\in\left(0,\frac{\overline{x}-\mu}{\overline{x}-\underline{x}}\right]}\frac{1+\underline{x}}{\rho(c)}-1=\left.\frac{\mu-\overline{x}+\underline{x}+1}{\rho(c)}+\overline{x}-\underline{x}-1\right|_{\rho(c)=\frac{\overline{x}-\mu}{\overline{x}-\underline{x}}}.\label{eq:fact}
\end{align}
For $c\in(\mu,\overline{x})$,
\begin{align}
\inf\limits_{\rho(c)\in\left(\frac{c-\mu}{c-\underline{x}},1\right)}\overline{\zeta}(c)
&=\min\left\{\inf\limits_{\rho(c)\in\left(\frac{c-\mu}{c-\underline{x}},\frac{\overline{x}-\mu}{\overline{x}-\underline{x}}\right]}\frac{1+\underline{x}}{\rho(c)}-1,\inf\limits_{\rho(c)\in\left(\frac{\overline{x}-\mu}{\overline{x}-\underline{x}},1\right)}\frac{\mu-\overline{x}+\underline{x}+1}{\rho(c)}+\overline{x}-\underline{x}-1\right\}\nonumber\\
&=\inf\limits_{\rho(c)\in\left(\frac{\overline{x}-\mu}{\overline{x}-\underline{x}},1\right)}\frac{\mu-\overline{x}+\underline{x}+1}{\rho(c)}+\overline{x}-\underline{x}-1\label{eq:twoinf2}\\
&=\begin{cases}
\frac{(1+\underline{x})(\overline{x}-\underline{x})}{\overline{x}-\mu}-1, &\mu\leq\overline{x}-\underline{x}-1,\\\mu, &\mu>\overline{x}-\underline{x}-1,
\end{cases}\label{eq:case1b}
\end{align}
where (\ref{eq:twoinf2}) is due to (\ref{eq:fact}).
One can readily prove (\ref{eq:Gdoubleunderlinec}) given (\ref{eq:case1a}) and (\ref{eq:case1b}).

\subsection{Proof of Proposition \ref{cor:upper2}}\label{sec:proofproposition4}

We shall only prove (\ref{eq:Gdoubleoverlinec}) since the proof of (\ref{eq:doubleoverlinec}) is similar to that of (\ref{eq:doubleunderlinec}).
Clearly, $r'(c)\geq\underline{\chi}(c)$ if and only if
\begin{align}
c\leq\begin{cases}
\sup\limits_{\rho(c)\in\left(0,\frac{\overline{x}-\mu}{\overline{x}-c}\right)}\underline{\zeta}(c), & c\in(\underline{x},\mu],\\
\sup\limits_{\rho(c)\in\left(\frac{c-\mu}{c-\underline{x}},1\right)}\underline{\zeta}(c), & c\in(\mu,\overline{x}),
\end{cases}\label{eq:hold}
\end{align}
where
\begin{align*}
\underline{\zeta}(c)=\begin{cases}
\frac{1+c}{\rho}-1,&c\in(\underline{x},\mu],\\
\frac{\mu+(1-\rho(c))(\rho(c)-c)}{\rho^2(c)},&c\in(\mu,\overline{x}).
\end{cases}
\end{align*}
For $c\in(\underline{x},\mu]$, 
\begin{align*}
\sup\limits_{\rho(c)\in\left(0,\frac{\overline{x}-\mu}{\overline{x}-c}\right)}\underline{\zeta}(c)=\infty
\end{align*}
and consequently (\ref{eq:hold}) trivially holds.
For $c\in(\mu,\overline{x})$, we have
\begin{align*}
\sup\limits_{\rho(c)\in\left(\frac{c-\mu}{c-\underline{x}},1\right)}\underline{\zeta}(c)=\begin{cases}
\frac{(c-\underline{x})(1+\underline{x})}{c-\mu}-1,& c\leq 2\underline{x}+1,\\
\frac{(1+c)^2}{4(c-\mu)}-1,& c\in(2\underline{x}+1,2\mu+1],\\
\mu, & c>2\mu+1,
\end{cases}
\end{align*}
which is a monotonically decreasing function of $c$. 
For $c>\mu$,
\begin{align*}
c=\frac{(c-\underline{x})(1+\underline{x})}{c-\mu}-1
\end{align*}
has a unique solution $c=c_1$, and 
\begin{align*}
c=\frac{(1+c)^2}{4(c-\mu)}-1
\end{align*}
has a unique solution $c=c_2$. Note that
\begin{align*}
\frac{(c-\underline{x})(1+\underline{x})}{c-\mu}\leq\frac{(1+c)^2}{4(c-\mu)},\quad c\in(\mu,2\mu+1].
\end{align*}
Therefore, $c_2\leq 2\underline{x}+1$ (i.e., $\mu\leq\frac{3}{2}\underline{x}+\frac{1}{2}$) implies $c_1\leq 2\underline{x}+1$. Now one can readily complete the proof of (\ref{eq:Gdoubleoverlinec}).


\section{Asymptotic Relationship Between $c^*$ and $\mu$ }\label{sec:example}


We shall focus on the reward function defined in (\ref{eq:awgn}) and provide a detailed analysis of $c^*$ for a few examples, with a particular interest in understanding how $c^*$ scales with $\mu$ as $\mu\downarrow0$ or $\mu\uparrow\infty$. In the sequel we write $c^*\sim_{0}\psi(\mu)$ ($c^*\sim_{\infty}\psi(\mu)$) to indicate that
$\lim_{\mu\downarrow0}\frac{c^*}{\psi(\mu)}=1$ ($\lim_{\mu\uparrow\infty}\frac{c^*}{\psi(\mu)}=1$). 


\subsection{Discrete Distribution}

\begin{itemize}
	\item Geometric distribution:
	\begin{align*}
	p_X(k)=(1-p)^kp,\quad k=0,1,\cdots, \quad p\in(0,1).
	\end{align*} 
	
	Note that $\mu=\frac{1-p}{p}$. Clearly,
	\begin{align*}
	c^*=\mu,\quad\mu\in(0,1],
	\end{align*}
	which implies $c^*\sim_{0}\mu$.

	For any $a>0$,
	\begin{align}
	\lim\limits_{\mu\uparrow\infty}\left(1+\frac{a\mu}{\log\mu}\right)\sum\limits_{k=0}^{\lfloor\frac{a\mu}{\log\mu}\rfloor}\frac{(1-p)^kp}{1+k}
	&=\lim\limits_{\mu\uparrow\infty}\left(1+\frac{a\mu}{\log\mu}\right)\sum\limits_{k=0}^{\lfloor\frac{a\mu}{\log\mu}\rfloor}\frac{\left(\frac{\mu}{1+\mu}\right)^k}{(1+\mu)(1+k)}\nonumber\\
	&=\lim\limits_{\mu\uparrow\infty}\left(1+\frac{a\mu}{\log\mu}\right)\sum\limits_{k=0}^{\lfloor\frac{a\mu}{\log\mu}\rfloor}\frac{1}{(1+\mu)(1+k)}\label{eq:uniform1}\\
	&=\lim\limits_{\mu\uparrow\infty}\left(1+\frac{a\mu}{\log\mu}\right)\frac{1}{1+\mu}\log\left(1+\left\lfloor\frac{a\mu}{\log\mu}\right\rfloor\right)\nonumber\\
	&=\lim\limits_{\mu\uparrow\infty}\frac{a}{\log\mu}\log\left(\frac{a\mu}{\log\mu}\right)\nonumber\\
	&=a,\nonumber
	\end{align}
	where (\ref{eq:uniform1}) is due to the fact that
	\begin{align*}
	1\geq\left(\frac{\mu}{1+\mu}\right)^k\geq\left(\frac{\mu}{1+\mu}\right)^{\frac{a\mu}{\log\mu}},\quad k=0,1,\cdots,\left\lfloor\frac{a\mu}{\log\mu}\right\rfloor,
	\end{align*}
	and 
	\begin{align*}
	\lim\limits_{\mu\uparrow\infty}\left(\frac{\mu}{1+\mu}\right)^{\frac{a\mu}{\log\mu}}=1.
	\end{align*}
	Therefore, we must have $c^*\sim_{\infty}\frac{\mu}{\log\mu}$.

	\item Poisson distribution:
	\begin{align*}
	p_X(k)=\frac{e^{-\lambda}\lambda^k}{k!},\quad k=0,1,\cdots,\quad\lambda>0.
	\end{align*}

	Note that $\mu=\mathbb{E}[(X-\mu)^2]=\lambda$.
	Clearly, 
\begin{align*}
c^*=e^{\mu}-1,\quad\mu\in(0,\log2],
\end{align*}
which implies $c^*\sim_{0}\mu$.

	It is shown in Appendix \ref{app:3terms} that for any $a>0$, 
	\begin{align}
		&\lim\limits_{\mu\uparrow\infty}(1+a\mu)\sum\limits_{k=\lceil\mu-\mu^{\frac{2}{3}}\rceil}^{\lfloor\mu+\mu^{\frac{2}{3}}\rfloor}\frac{e^{-\lambda}\lambda^k}{(1+k)(k!)}=a,\label{eq:term1}\\
		&\lim\limits_{\mu\uparrow\infty}(1+a\mu)\sum\limits_{k=\lfloor\mu+\mu^{\frac{2}{3}}\rfloor+1}^\infty\frac{e^{-\lambda}\lambda^k}{(1+k)(k!)}=0,\label{eq:term2}\\
		&\lim\limits_{\mu\uparrow\infty}(1+a\mu)\sum\limits_{k=0}^{\lceil\mu-\mu^{\frac{2}{3}}\rceil-1}\frac{e^{-\lambda}\lambda^k}{(1+k)(k!)}=0.\label{eq:term3}
	\end{align}

	
	Therefore, we have
	\begin{align*}
	\lim\limits_{\mu\uparrow\infty}(1+a\mu)\sum\limits_{k=0}^{\lfloor a\mu\rfloor}\frac{e^{-\lambda}\lambda^k}{(1+k)(k!)}=
	\begin{cases}
	0,&a<1,\\
	a,&a>1,
	\end{cases}
	\end{align*}
	which implies
$c^*\sim_{\infty}\mu$.

\end{itemize}

\begin{figure*}[htbp]
	\centerline{\includegraphics[width=18cm]{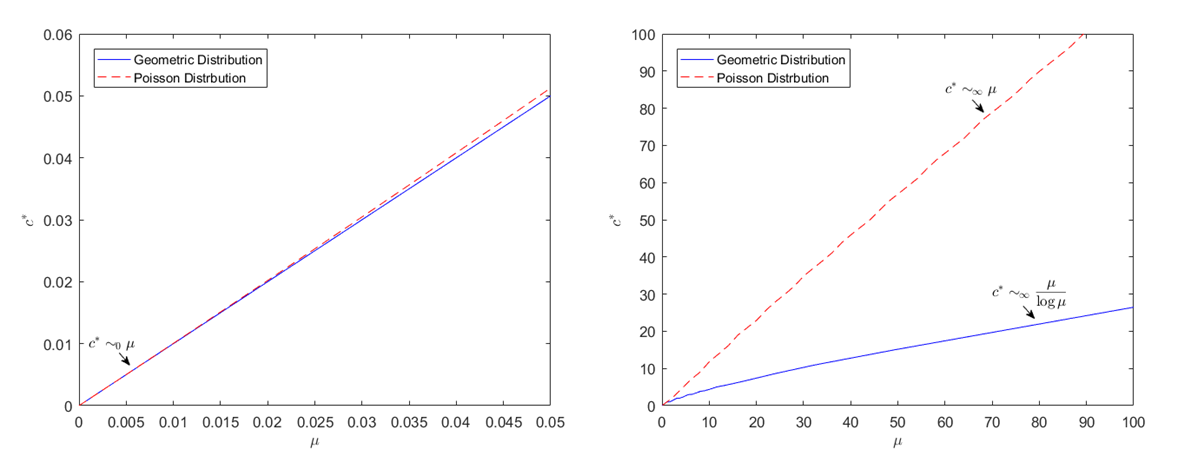}}
	\caption{The relationship between $c^*$ and $\mu$ for some discrete distributions}
	\label{fig:discrete}
\end{figure*}

We plot $c^*$ against $\mu$ in Fig. \ref{fig:discrete} for the geometric distribution and the Poisson distribution, which confirms our asymptotic analysis.


\subsection{Continuous Distribution}

\begin{itemize}
	\item Uniform distribution:
	\begin{align*}
	f_X(x)=\begin{cases}
	\frac{1}{\omega}, & x\in[0,\omega],\\
	0, & x\notin[0,\omega],
	\end{cases}\quad\omega>0.
		\end{align*}
	
We can write (\ref{eq:contin}) equivalently as
\begin{align*}
\frac{1+c^*}{\omega}\log(1+c^*)=1.
\end{align*}
Note that $\mu=\frac{\omega}{2}$. For any $a>0$,
\begin{align*}
\lim\limits_{\mu\downarrow 0}\frac{1+a\mu}{\omega}\log\left(1+a\mu\right)=\lim\limits_{\mu\downarrow 0}\frac{1+a\mu}{2\mu}\log\left(1+a\mu\right)=\frac{a}{2}.
\end{align*}
Therefore, we must have
$c^*\sim_{0} 2\mu$.

For any $a>0$, 
\begin{align*}
\lim\limits_{\mu\uparrow\infty}\frac{1+\frac{a\mu}{\log\mu}}{\omega}\log\left(1+\frac{a\mu}{\log\mu}\right)=\lim\limits_{\mu\uparrow\infty}\frac{1+\frac{a\mu}{\log\mu}}{2\mu}\log\left(1+\frac{a\mu}{\log\mu}\right)=\lim\limits_{\mu\uparrow\infty}\frac{a}{2\log\mu}\log\left(\frac{a\mu}{\log\mu}\right)=\frac{a}{2}.
\end{align*}
Therefore, we must have
	$c^*\sim_{\infty} \frac{2\mu}{\log\mu}$.

	\item Exponential distribution:
	\begin{align*}
	f_X(x)=\begin{cases}
	\eta e^{-\eta x},& x\geq 0,\\
	0,& x<0,
	\end{cases}\quad\eta>0.
	\end{align*}
	
	We can write (\ref{eq:contin}) equivalently as
	\begin{align*}
	(1+c^*)\int_0^{c^*}\frac{\eta e^{-\eta x}}{1+x}\mathrm{d}x=1.
	\end{align*}	
	Note that $\mu=\frac{1}{\eta}$. For any $a>0$,
	\begin{align}
	(1-a\mu\log\mu)\int_{0}^{-a\mu\log\mu}\frac{\eta e^{-\eta x}}{1+x}\mathrm{d}x
	&=(1-a\mu\log\mu)\int_{0}^{-a\mu\log\mu}\frac{e^{-\frac{x}{\mu}}}{\mu(1+x)}\mathrm{d}x\nonumber\\
	&=(1-a\mu\log\mu)\int_{0}^{-a\mu\log\mu}\frac{e^{-\frac{x}{\mu}}}{\mu}(1-x+o_{x\downarrow 0}(x))\mathrm{d}x.\label{eq:exp1+2}
	\end{align}
	It can be verified that
	\begin{align}
	\int_{0}^{-a\mu\log\mu}\frac{e^{-\frac{x}{\mu}}}{\mu}\mathrm{d}x=1-\mu^a,\label{eq:exp1}
	\end{align}
	and
	\begin{align}
	\int_{0}^{-a\mu\log\mu}\frac{xe^{-\frac{x}{\mu}}}{\mu}\mathrm{d}x=a\mu^{a+1}\log\mu-\mu^{a+1}+\mu.\label{eq:exp2}
	\end{align}
	Substituting (\ref{eq:exp1}) and (\ref{eq:exp2}) into (\ref{eq:exp1+2}) gives
	\begin{align*}
	(1-a\mu\log\mu)\int_{0}^{-a\mu\log\mu}\frac{\eta e^{-\eta x}}{1+x}\mathrm{d}x=1-a\mu\log\mu-\mu^a+o_{\mu\downarrow 0}(\mu\log\mu).
	\end{align*}
	When $\mu$ is sufficiently close to 0, 
	\begin{align*}
	1-a\mu\log\mu-\mu^a+o_{\mu\downarrow 0}(\mu\log\mu)\begin{cases}
	<1, &a<1,\\
	>1, &a>1.
	\end{cases}
	\end{align*}
	Therefore, we must have $c^*\sim_0-\mu\log\mu$.

	For any $a>0$, 
	\begin{align}
	\lim\limits_{\mu\uparrow\infty}\left(1+\frac{a\mu}{\log\mu}\right)\int_0^{\frac{a\mu}{\log\mu}}\frac{\eta e^{-\eta x}}{1+x}\mathrm{d}x
	&=\lim\limits_{\mu\uparrow\infty}\left(1+\frac{a\mu}{\log\mu}\right)\int_0^{\frac{a\mu}{\log\mu}}\frac{e^{-\frac{x}{\mu}}}{\mu(1+x)}\mathrm{d}x\nonumber\\
		&=\lim\limits_{\mu\uparrow\infty}\left(1+\frac{a\mu}{\log\mu}\right)\int_0^{\frac{a\mu}{\log\mu}}\frac{1}{\mu(1+x)}\mathrm{d}x\label{eq:uniform2}\\
		&=\lim\limits_{\mu\uparrow\infty}\left(1+\frac{a\mu}{\log\mu}\right)\frac{1}{\mu}\log\left(1+\frac{a\mu}{\log\mu}\right)\nonumber\\
		&=\lim\limits_{\mu\uparrow\infty}\frac{a}{\log\mu}\log\left(\frac{a\mu}{\log\mu}\right)\nonumber\\
		&=a,\nonumber
	\end{align}
	where (\ref{eq:uniform2}) is due to the fact that
	\begin{align*}
	1\geq e^{-\frac{x}{\mu}}\geq e^{-\frac{a}{\log\mu}}, \quad x\in\left[0,\frac{a\mu}{\log\mu}\right],
	\end{align*}
	and
	\begin{align*}
	\lim\limits_{\mu\uparrow\infty}e^{-\frac{a}{\log\mu}}=1.
	\end{align*}
	Therefore, we must have
	$c^*\sim_\infty\frac{\mu}{\log\mu}$.

	\item Rayleigh distribution:
	\begin{align*}
	f_X(x)=\begin{cases}
	\frac{x}{\theta}e^{-\frac{x^2}{2\theta}}, &x\geq 0,\\
	0,&x<0,
	\end{cases}\quad\theta>0.
	\end{align*}

	We can write (\ref{eq:contin}) equivalently as
	\begin{align*}
(1+c^*)\int_0^{c^*}\frac{xe^{-\frac{x^2}{2\theta}}}{\theta(1+x)}\mathrm{d}x=1.
\end{align*}
Note that $\mu=\sqrt{\frac{\pi\theta}{2}}$. For any $a>0$,
\begin{align}
&(1+a\mu\sqrt{-\log\mu})\int_0^{a\mu\sqrt{-\log\mu}}\frac{xe^{-\frac{x^2}{2\theta}}}{\theta(1+x)}\mathrm{d}x\nonumber\\
&=(1+a\mu\sqrt{-\log\mu})\int_0^{a\mu\sqrt{-\log\mu}}\frac{\pi xe^{-\frac{\pi x^2}{4\mu^2}}}{2\mu^2(1+x)}\mathrm{d}x\nonumber\\
&=(1+a\mu\sqrt{-\log\mu})\int_0^{a\mu\sqrt{-\log\mu}}\frac{\pi e^{-\frac{\pi x^2}{4\mu^2}}}{2\mu^2}(x-x^2+o_{x\downarrow 0}(x^2))\mathrm{d}x.\label{eq:rayleigh1+2}
\end{align}
It can be verified that
\begin{align}
\int_0^{a\mu\sqrt{-\log\mu}}\frac{\pi xe^{-\frac{\pi x^2}{4\mu^2}}}{2\mu^2}\mathrm{d}x=1-\mu^{\frac{\pi a^2}{4}},\label{eq:rayleigh1}
\end{align}
and
\begin{align}
\int_0^{a\mu\sqrt{-\log\mu}}\frac{\pi x^2e^{-\frac{\pi x^2}{4\mu^2}}}{2\mu^2}\mathrm{d}x=-a\mu^{\frac{\pi a^2}{4}+1}\sqrt{-\log\mu}+\mu\int_0^{a\sqrt{-\log\mu}}e^{-\frac{\pi y^2}{4}}\mathrm{d}y.\label{eq:rayleigh2}
\end{align}
Substituting (\ref{eq:rayleigh1}) and (\ref{eq:rayleigh2}) into (\ref{eq:rayleigh1+2}) gives
\begin{align*}
(1+a\mu\sqrt{-\log\mu})\int_0^{a\mu\sqrt{-\log\mu}}\frac{xe^{-\frac{x^2}{2\theta}}}{\theta(1+x)}\mathrm{d}x\nonumber=1+a\mu\sqrt{-\log\mu}-\mu^{\frac{\pi a^2}{4}}+o_{\mu\downarrow 0}(\mu\sqrt{-\log\mu}).
\end{align*}
When $\mu$ is sufficiently close to 0, 
\begin{align*}
1+a\mu\sqrt{-\log\mu}-\mu^{\frac{\pi a^2}{4}}+o_{\mu\downarrow 0}(\mu\sqrt{-\log\mu})\begin{cases}
<1, &a<\frac{2}{\sqrt{\pi}},\\
>1, &a>\frac{2}{\sqrt{\pi}}.
\end{cases}
\end{align*}
Therefore, we must have $c^*\sim_0\frac{2}{\sqrt{\pi}}\mu\sqrt{-\log\mu}$.

For any $a>0$,
\begin{align}
&\lim\limits_{\mu\uparrow\infty}(1+a\mu)\int_0^{a\mu}\frac{xe^{-\frac{x^2}{2\theta}}}{\theta(1+x)}\mathrm{d}x\nonumber\\
&=\lim\limits_{\mu\uparrow\infty}(1+a\mu)\int_0^{a\mu}\frac{\pi xe^{-\frac{\pi x^2}{4\mu^2}}}{2\mu^2(1+x)}\mathrm{d}x\nonumber\\
&=\lim\limits_{\mu\uparrow\infty}(1+a\mu)\int_{0}^{\log\mu}\frac{\pi xe^{-\frac{\pi x^2}{4\mu^2}}}{2\mu^2(1+x)}\mathrm{d}x+\lim\limits_{\mu\uparrow\infty}(1+a\mu)\int_{\log\mu}^{a\mu}\frac{\pi xe^{-\frac{\pi x^2}{4\mu^2}}}{2\mu^2(1+x)}\mathrm{d}x.\label{eq:int1+int2}
\end{align}
It can be verified that
\begin{align*}
0\leq\lim\limits_{\mu\uparrow\infty}(1+a\mu)\int_{0}^{\log\mu}\frac{\pi xe^{-\frac{\pi x^2}{4\mu^2}}}{2\mu^2(1+x)}\mathrm{d}x\leq\lim\limits_{\mu\uparrow\infty}(1+a\mu)\int_{0}^{\log\mu}\frac{\pi}{2\mu^2}\mathrm{d}x=\lim\limits_{\mu\uparrow\infty}\frac{\pi(1+a\mu)\log\mu}{2\mu^2}=0,
\end{align*}
which implies
\begin{align}
\lim\limits_{\mu\uparrow\infty}(1+a\mu)\int_{0}^{\log\mu}\frac{\pi xe^{-\frac{\pi x^2}{4\mu^2}}}{2\mu^2(1+x)}\mathrm{d}x=0.\label{eq:int1}
\end{align}
Moreover, 
\begin{align}
\lim\limits_{\mu\uparrow\infty}(1+a\mu)\int_{\log\mu}^{a\mu}\frac{\pi xe^{-\frac{\pi x^2}{4\mu^2}}}{2\mu^2(1+x)}\mathrm{d}x
&=\lim\limits_{\mu\uparrow\infty}(1+a\mu)\int_{\log\mu}^{a\mu}\frac{\pi e^{-\frac{\pi x^2}{4\mu^2}}}{2\mu^2}\mathrm{d}x\label{eq:uniform3}\\
&=\lim\limits_{\mu\uparrow\infty}(1+a\mu)\int_{\frac{\log\mu}{\mu}}^a\frac{\pi e^{-\frac{\pi y^2}{4}}}{2\mu}\mathrm{d}y\nonumber\\
&=\frac{\pi a}{2}\int_0^a e^{-\frac{\pi y^2}{4}}\mathrm{d}y,\label{eq:int2}
\end{align}
where (\ref{eq:uniform3}) is due to the fact that
\begin{align*}
\frac{\log\mu}{1+\log\mu}\leq\frac{x}{1+x}\leq\frac{a\mu}{1+a\mu},\quad x\in[\log\mu,a\mu],
\end{align*}
and
\begin{align*}
\lim\limits_{\mu\uparrow\infty}\frac{\log\mu}{1+\log\mu}=\lim\limits_{\mu\uparrow\infty}\frac{a\mu}{1+a\mu}=1.
\end{align*}
Substituting (\ref{eq:int1}) and (\ref{eq:int2}) into (\ref{eq:int1+int2}) gives
\begin{align*}
\lim\limits_{\mu\uparrow\infty}(1+a\mu)\int_0^{a\mu}\frac{xe^{-\frac{x^2}{2\theta}}}{\theta(1+x)}\mathrm{d}x=\frac{\pi a}{2}\int_0^a e^{-\frac{\pi y^2}{4}}\mathrm{d}y.
\end{align*}
Therefore, we must have
$c^*\sim_{\infty} a^*\mu$,
where $a^*\approx0.875$ is the unique positive number satisfying
\begin{align*}
\frac{\pi a^*}{2}\int_0^{a^*} e^{-\frac{\pi y^2}{4}}\mathrm{d}y=1.
\end{align*}

\end{itemize}

\begin{figure*}[htbp]
	\centerline{\includegraphics[width=18cm]{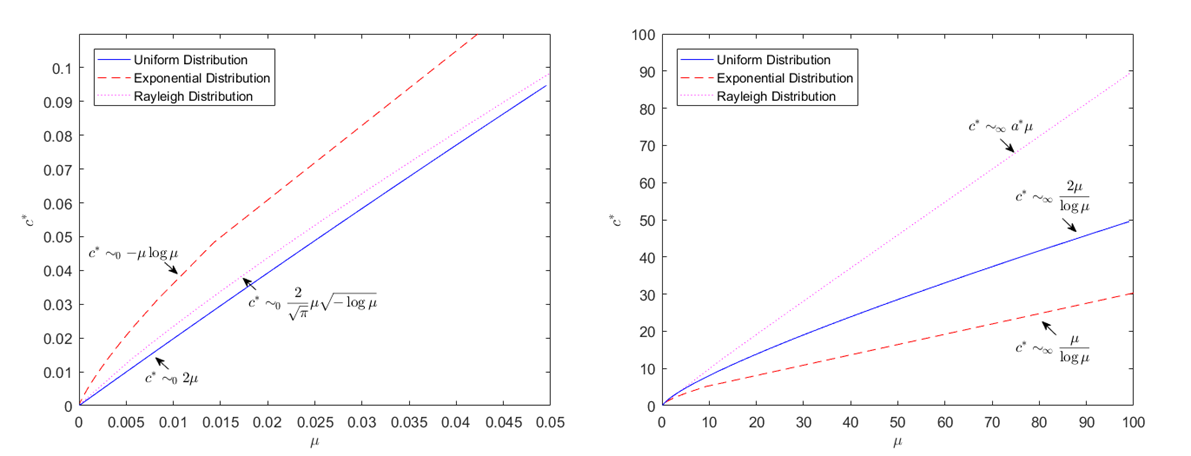}}
	\caption{The relationship between $c^*$ and $\mu$ for some continuous distributions.}
	\label{fig:continuous}
\end{figure*}

We plot $c^*$ against $\mu$ in Fig. \ref{fig:continuous} for the uniform distribution, the exponential distribution, and the Rayleigh distribution, which confirms our asymptotic analysis.

\section{Conclusion}\label{sec:conclusion}

We have studied the problem of online power control for battery limited energy harvesting communications. The main finding of this work is that the greedy policy achieves the maximum throughput if and only if the battery capacity is below a certain threshold. It is worth noting that this threshold depends on the distribution of the energy arrival process although the greedy policy itself does not. In fact, 
there does not exist a positive threshold on the battery capacity below which the greedy policy (or any other universal policy) is throughput-optimal for all energy arrival processes. Nevertheless, as shown in \cite{SO16}, it is possible to define certain weakened notion of universality and optimality, and construct the associated online power control poicy. Further progress along this line of research can be found in \cite{YC19}.



\appendices

\section{Proof of a Statement in Remark  \ref{remark:overlinec}}\label{app:overlinec}

We assume $\overline{x}=\infty$ throughout this proof.

First consider the case $r'(\mu-\epsilon)=r'(\overline{x})$ for some $\epsilon>0$, which implies
\begin{align}
r'(x)=r'(\overline{x}),\quad x\geq\mu-\epsilon.\label{eq:neighbor}
\end{align} 
Note that for $c\geq\mu$,
\begin{align}
&r'(c)=r'(\overline{x}),\label{eq:inf1}\\
&\rho(c)\underline{r'}_{[\underline{x},c]}(\overline{\xi})\leq r'\left(\frac{\mu-(1-\rho(c))c}{\rho(c)}\right).\label{eq:inf2}
\end{align}
Moreover, in view of (\ref{eq:neighbor}) and the fact that
\begin{align*}
\lim\limits_{c\uparrow\overline{x}}\frac{\mu-(1-\rho(c))c}{\rho(c)}=\mu,
\end{align*}
we have 
\begin{align}
r'\left(\frac{\mu-(1-\rho(c))c}{\rho(c)}\right)=r'(\overline{x})\label{eq:inf3}
\end{align}
for all sufficiently large $c$. Combining (\ref{eq:inf1}), (\ref{eq:inf2}), and (\ref{eq:inf3}) proves $\overline{c}=\infty$.

Next consider the case $r'(\mu)>r'(\overline{x})$. There must exist $\epsilon>0$ such that $r'(\mu+\epsilon)>r'(\overline{x})$. For $c>\underline{x}$ and $x\in[\underline{x},\min\{\mu,c\}]$, it is easy to establish the following uniform lower bound:	
\begin{align*}
\underline{r'}_{[\underline{x},c]}(x)\geq\min\left\{\frac{\epsilon}{\mu+\epsilon-\underline{x}}r'(\mu)+\frac{\mu-\underline{x}}{\mu+\epsilon-\underline{x}}r'(\overline{x}), r'(\mu+\epsilon)\right\}.
\end{align*}
Clearly, we have $\overline{\xi}\in[\underline{x},\min\{\mu,c\}]$ for $c>\underline{x}$.
Now it can be readily verified that
\begin{align*}
\lim\limits_{c\uparrow\overline{x}}\rho(c)\underline{r'}_{[\underline{x},c]}(\overline{\xi})\geq\min\left\{\frac{\epsilon}{\mu+\epsilon-\underline{x}}r'(\mu)+\frac{\mu-\underline{x}}{\mu+\epsilon-\underline{x}}r'(\overline{x}), r'(\mu+\epsilon)\right\}>r'(\overline{x}),
\end{align*}
which implies $\overline{x}<\infty$.

\section{Proof of (\ref{eq:left}) and (\ref{eq:right})}\label{app:leftright}

Note that
\begin{align*}
&\phi(g)-\phi(g-\epsilon)\\
&=r(g)-r(g-\epsilon)+\rho(c-b+g)\mathbb{E}[r(b-g+X)|X<c-b+g]\\
&\quad-\rho(c-b+g-\epsilon)\mathbb{E}[r(b-g+\epsilon+X)|X<c-b+g-\epsilon]-\mathbb{P}(c-b+g-\epsilon\leq X<c-b+g)r(c)\\
&=r(g)-r(g-\epsilon)+\rho(c-b+g)\mathbb{E}[r(b-g+X)|X<c-b+g]\\
&\quad-\rho(c-b+g)\mathbb{E}[r(b-g+\epsilon+X)|X<c-b+g]+\rho(c-b+g)\mathbb{E}[r(b-g+\epsilon+X)|X<c-b+g]\\
&\quad-\rho(c-b+g-\epsilon)\mathbb{E}[r(b-g+\epsilon+X)|X<c-b+g-\epsilon]-\mathbb{P}(c-b+g-\epsilon\leq X<c-b+g)r(c)\\
&=r(g)-r(g-\epsilon)+\rho(c-b+g)\mathbb{E}[r(b-g+X)-r(b-g+\epsilon+X)|X<c-b+g]\\
&\quad+\mathbb{P}(c-b+g-\epsilon\leq X<c-b+g)\mathbb{E}[r(b-g+\epsilon+X)-r(c)|c-b+g-\epsilon\leq X<c-b+g].
\end{align*}
Therefore,
\begin{align}
&\lim\limits_{\epsilon\downarrow 0}\frac{1}{\epsilon}(\phi(g)-\phi(g-\epsilon))\nonumber\\
&=\lim\limits_{\epsilon\downarrow 0}\frac{1}{\epsilon}(r(g)-r(g-\epsilon))+\lim\limits_{\epsilon\downarrow 0}\frac{1}{\epsilon}\rho(c-b+g)\mathbb{E}[r(b-g+X)-r(b-g+\epsilon+X)|X<c-b+g]\nonumber\\
&\quad+\lim\limits_{\epsilon\downarrow 0}\frac{1}{\epsilon}\mathbb{P}(c-b+g-\epsilon\leq X<c-b+g)\mathbb{E}[r(b-g+\epsilon+X)-r(c)|c-b+g-\epsilon\leq X<c-b+g].\label{eq:suma}
\end{align}
Clearly, we have
\begin{align}
\lim\limits_{\epsilon\downarrow 0}\frac{1}{\epsilon}(r(g)-r(g-\epsilon))=r'(g).\label{eq:limita1}
\end{align}
Moreover, in light of \cite[Theorem 9.1, p.~481]{Durrett95},
\begin{align}
&\lim\limits_{\epsilon\downarrow 0}\frac{1}{\epsilon}\rho(c-b+g)\mathbb{E}[r(b-g+X)-r(b-g+\epsilon+X)|X<c-b+g]\nonumber\\
&=-\rho(c-b+g)\mathbb{E}[r(b-g+X)|X<c-b+g].\label{eq:limita2}
\end{align}
It can also be verified that
\begin{align*}
0&\leq\liminf\limits_{\epsilon\downarrow 0}\frac{1}{\epsilon}\mathbb{P}(c-b+g-\epsilon\leq X<c-b+g)\mathbb{E}[r(b-g+\epsilon+X)-r(c)|c-b+g-\epsilon\leq X<c-b+g]\nonumber\\
&\leq\limsup\limits_{\epsilon\downarrow 0}\frac{1}{\epsilon}\mathbb{P}(c-b+g-\epsilon\leq X<c-b+g)\mathbb{E}[r(b-g+\epsilon+X)-r(c)|c-b+g-\epsilon\leq X<c-b+g]\nonumber\\
&\leq\limsup\limits_{\epsilon\downarrow 0}\frac{1}{\epsilon}\mathbb{P}(c-b+g-\epsilon\leq X<c-b+g)\mathbb{E}[r'(c)(b-g+\epsilon+X-c)|c-b+g-\epsilon\leq X<c-b+g]\nonumber\\
&\leq\limsup\limits_{\epsilon\downarrow 0}\mathbb{P}(c-b+g-\epsilon\leq X<c-b+g)r'(c)\nonumber\\
&=0,
\end{align*}
which implies
\begin{align}
&\lim\limits_{\epsilon\downarrow 0}\frac{1}{\epsilon}\mathbb{P}(c-b+g-\epsilon\leq X<c-b+g)\mathbb{E}[r(b-g+\epsilon+X)-r(c)|c-b+g-\epsilon\leq X<c-b+g]\nonumber\\
&=0.\label{eq:limita3}
\end{align}
Substituting (\ref{eq:limita1}), (\ref{eq:limita2}), and (\ref{eq:limita3}) into (\ref{eq:suma}) proves (\ref{eq:left}).

Note that
\begin{align*}
&\phi(g+\epsilon)-\phi(g)\\
&=r(g+\epsilon)-r(g)+\rho(c-b+g+\epsilon)\mathbb{E}[r(b-g-\epsilon+X)|X<c-b+g+\epsilon]\\
&\quad-\rho(c-b+g)\mathbb{E}[r(b-g+X)|X<c-b+g]-\mathbb{P}(c-b+g\leq X<c-b+g+\epsilon)r(c)\\
&=r(g+\epsilon)-r(g)+\rho(c-b+g+\epsilon)\mathbb{E}[r(b-g-\epsilon+X)|X<c-b+g+\epsilon]\\
&\quad-\rho(c-b+g)\mathbb{E}[r(b-g-\epsilon+X)|X<c-b+g]+\rho(c-b+g)\mathbb{E}[r(b-g-\epsilon+X)|X<c-b+g]\\
&\quad-\rho(c-b+g)\mathbb{E}[r(b-g+X)|X<c-b+g]-\mathbb{P}(c-b+g\leq X<c-b+g+\epsilon)r(c)\\
&=r(g+\epsilon)-r(g)+\rho(c-b+g)\mathbb{E}[r(b-g-\epsilon+X)-r(b-g+X)|X<c-b+g]\\
&\quad+\mathbb{P}(c-b+g\leq X<c-b+g+\epsilon)\mathbb{E}[r(b-g-\epsilon+X)-r(c)|c-b+g\leq X<c-b+g+\epsilon]\\
&=r(g+\epsilon)-r(g)+\rho(c-b+g)\mathbb{E}[r(b-g-\epsilon+X)-r(b-g+X)|X<c-b+g]\\
&\quad+\mathbb{P}(c-b+g< X<c-b+g+\epsilon)\mathbb{E}[r(b-g-\epsilon+X)-r(c)|c-b+g< X<c-b+g+\epsilon]\\
&\quad+\mathbb{P}(X=c-b+g)(r(c-\epsilon)-r(c)).
\end{align*}
Therefore,
\begin{align}
&\lim\limits_{\epsilon\downarrow 0}\frac{1}{\epsilon}(\phi(g+\epsilon)-\phi(g))\nonumber\\
&=\lim\limits_{\epsilon\downarrow 0}\frac{1}{\epsilon}(r(g+\epsilon)-r(g))+\lim\limits_{\epsilon\downarrow 0}\frac{1}{\epsilon}\rho(c-b+g)\mathbb{E}[r(b-g-\epsilon+X)-r(b-g+X)|X<c-b+g]\nonumber\\
&\quad+\lim\limits_{\epsilon\downarrow 0}\frac{1}{\epsilon}\mathbb{P}(c-b+g< X<c-b+g+\epsilon)\mathbb{E}[r(b-g-\epsilon+X)-r(c)|c-b+g< X<c-b+g+\epsilon]\nonumber\\
&\quad+\lim\limits_{\epsilon\downarrow 0}\frac{1}{\epsilon}\mathbb{P}(X=c-b+g)(r(c-\epsilon)-r(c)).\label{eq:sumb}
\end{align}
Similarly to (\ref{eq:limita1}), (\ref{eq:limita2}), and (\ref{eq:limita3}), we have
\begin{align}
&\lim\limits_{\epsilon\downarrow 0}\frac{1}{\epsilon}(r(g+\epsilon)-r(g))=r'(g),\label{eq:limitb1}\\
&\lim\limits_{\epsilon\downarrow 0}\frac{1}{\epsilon}\rho(c-b+g)\mathbb{E}[r(b-g-\epsilon+X)-r(b-g+X)|X<c-b+g]\nonumber\\
&=-\rho(c-b+g)\mathbb{E}[r'(b-g+X)|X<c-b+g],\label{eq:limitb2}\\
&\lim\limits_{\epsilon\downarrow 0}\frac{1}{\epsilon}\mathbb{P}(c-b+g<X< c-b+g+\epsilon)\mathbb{E}[r(b-g-\epsilon+X)-r(c)|c-b+g< X<c-b+g+\epsilon]\nonumber\\
&=0.\label{eq:limitb3}
\end{align}
Moreover,
\begin{align}
\lim\limits_{\epsilon\downarrow 0}\frac{1}{\epsilon}\mathbb{P}(X=c-b+g)(r(c-\epsilon)-r(c))=-\mathbb{P}(X=c-b+g)r'(c).\label{eq:limitb4}
\end{align}
Substituting (\ref{eq:limitb1}), (\ref{eq:limitb2}), (\ref{eq:limitb3}), and (\ref{eq:limitb4}) into (\ref{eq:sumb}) proves (\ref{eq:right}).



\section{Proof of (\ref{eq:term1}), (\ref{eq:term2}), and (\ref{eq:term3})}\label{app:3terms}

We have
\begin{align}
\mathbb{P}(\lceil\mu-\mu^{\frac{2}{3}}\rceil\leq X\leq\lfloor\mu+\mu^{\frac{2}{3}}\rfloor)
&\geq
\mathbb{P}(|X-\mu|<\mu^{\frac{2}{3}})\nonumber\\	&=1-\mathbb{P}(|X-\mu|\geq\mu^{\frac{2}{3}})\nonumber\\
&\geq 1-\frac{\mathbb{E}[(X-\mu)^2]}{\mu^{\frac{4}{3}}}\label{eq:invokechebyshev}\\
&=1-\mu^{-\frac{1}{3}},\nonumber
\end{align}
where (\ref{eq:invokechebyshev}) is due to Chebyshev's inequality. Therefore,
\begin{align*}
\lim\limits_{\mu\uparrow \infty}\mathbb{P}(\lceil\mu-\mu^{\frac{2}{3}}\rceil\leq X\leq\lfloor\mu+\mu^{\frac{2}{3}}\rfloor)=1.
\end{align*}	

It can be verified that 
\begin{align}
\lim\limits_{\mu\uparrow\infty}(1+a\mu)\sum\limits_{k=\lceil\mu-\mu^{\frac{2}{3}}\rceil}^{\lfloor\mu+\mu^{\frac{2}{3}}\rfloor}\frac{e^{-\lambda}\lambda^k}{(1+k)(k!)}
&=\lim\limits_{\mu\uparrow\infty}a\sum\limits_{k=\lceil\mu-\mu^{\frac{2}{3}}\rceil}^{\lfloor\mu+\mu^{\frac{2}{3}}\rfloor}\frac{e^{-\lambda}\lambda^k}{k!}\label{eq:chebyshev1}\\
&=\lim\limits_{\mu\uparrow\infty}a\mathbb{P}(\lceil\mu-\mu^{\frac{2}{3}}\rceil\leq X\leq\lfloor\mu+\mu^{\frac{2}{3}}\rfloor)\nonumber\\
&=a,\nonumber
\end{align}
where (\ref{eq:chebyshev1}) is due to the fact that
\begin{align*}
\frac{1+a\mu}{1+\lfloor\mu+\mu^{\frac{2}{3}}\rfloor}\leq\frac{1+a\mu}{1+k}\leq\frac{1+a\mu}{1+\lceil\mu-\mu^{\frac{2}{3}}\rceil},\quad k=\lceil\mu-\mu^{\frac{2}{3}}\rceil,\cdots,\lfloor\mu+\mu^{\frac{2}{3}}\rfloor,
\end{align*}
and
\begin{align*}
\lim\limits_{\mu\uparrow\infty}\frac{1+a\mu}{1+\lfloor\mu+\mu^{\frac{2}{3}}\rfloor}=\lim\limits_{\mu\uparrow\infty}\frac{1+a\mu}{1+\lceil\mu-\mu^{\frac{2}{3}}\rceil}=a.
\end{align*}
This proves (\ref{eq:term1}).

It can also be verified that
\begin{align}
\lim\limits_{\mu\uparrow\infty}(1+a\mu)\sum\limits_{k=\lfloor\mu+\mu^{\frac{2}{3}}\rfloor+1}^\infty\frac{e^{-\lambda}\lambda^k}{(1+k)(k!)}&\leq \lim\limits_{\mu\uparrow\infty}a\sum\limits_{k=\lfloor\mu+\mu^{\frac{2}{3}}\rfloor+1}^\infty\frac{e^{-\lambda}\lambda^k}{k!}\label{eq:chebyshev2}\\
&=\lim\limits_{\mu\uparrow\infty}a\mathbb{P}(X\geq\lfloor\mu+\mu^{\frac{2}{3}}\rfloor+1)\nonumber\\
&\leq\lim\limits_{\mu\uparrow\infty}a(1-\mathbb{P}(\lceil\mu-\mu^{\frac{2}{3}}\rceil\leq X\leq\lfloor\mu+\mu^{\frac{2}{3}}\rfloor))\nonumber\\
&=0,\nonumber
\end{align}
where is (\ref{eq:chebyshev2}) due to the fact that
\begin{align*}
\frac{1+a\mu}{1+k}\leq\frac{1+a\mu}{\lfloor\mu+\mu^{\frac{2}{3}}\rfloor+2},\quad k\geq\lfloor\mu+\mu^{\frac{2}{3}}\rfloor+1,
\end{align*}
and
\begin{align*}
\lim\limits_{\mu\uparrow\infty}\frac{1+a\mu}{\lfloor\mu+\mu^{\frac{2}{3}}\rfloor+2}=a.
\end{align*}
This proves (\ref{eq:term2}).

Finally, note that
\begin{align}
\lim\limits_{\mu\uparrow\infty}(1+a\mu)\sum\limits_{k=0}^{\lceil\mu-\mu^{\frac{2}{3}}\rceil-1}\frac{e^{-\lambda}\lambda^k}{(1+k)(k!)}
&=\lim\limits_{\mu\uparrow\infty}(1+a\mu)\sum\limits_{k=0}^{\lceil\mu-\mu^{\frac{2}{3}}\rceil-1}\frac{e^{-\mu}\mu^k}{(1+k)(k!)}\nonumber\\
&\leq	\lim\limits_{\mu\uparrow\infty}(1+a\mu)\sum\limits_{k=0}^{\lfloor\mu-\mu^{\frac{2}{3}}\rfloor-1}\frac{e^{-\mu}\mu^k}{k!}\nonumber\\
&\leq\lim\limits_{\mu\uparrow\infty}(1+a\mu)\lfloor\mu-\mu^{\frac{2}{3}}\rfloor\frac{e^{-\mu}\mu^{\lfloor\mu-\mu^{\frac{2}{3}}\rfloor}}{\lfloor\mu-\mu^{\frac{2}{3}}\rfloor!},\label{eq:mode}
\end{align}
where (\ref{eq:mode}) is due to the fact that
\begin{align*}
\frac{\mu^k}{k!}\leq\frac{\mu^{\lfloor\mu-\mu^{\frac{2}{3}}\rfloor}}{\lfloor\mu-\mu^{\frac{2}{3}}\rfloor!},\quad k=0,1,\cdots,\lfloor\mu-\mu^{\frac{2}{3}}\rfloor-1.
\end{align*}
Let $\delta\triangleq\frac{\mu-\lfloor\mu-\mu^{\frac{2}{3}}\rfloor}{\mu}$. We have
\begin{align}
\lim\limits_{\mu\uparrow\infty}(1+a\mu)\lfloor\mu-\mu^{\frac{2}{3}}\rfloor\frac{e^{-\mu}\mu^{\lfloor\mu-\mu^{\frac{2}{3}}\rfloor}}{\lfloor\mu-\mu^{\frac{2}{3}}\rfloor!}&=\lim\limits_{\mu\uparrow\infty}(1+a\mu)\mu(1-\delta)\frac{e^{-\mu}\mu^{\mu(1-\delta)}}{(\mu(1-\delta))!}\nonumber\\
&=\lim\limits_{\mu\uparrow\infty}(1+a\mu)\mu(1-\delta)\frac{e^{-\mu\delta}(1-\delta)^{-\mu(1-\delta)-\frac{1}{2}}}{\sqrt{2\pi\mu}},\label{eq:stirling}
\end{align}
where   (\ref{eq:stirling}) is a consequence of Stirling's approximation $(\mu(1-\delta))!\sim_{\infty}\sqrt{2\pi\mu(1-\delta)}e^{-\mu(1-\delta)}(\mu(1-\delta))^{\mu(1-\delta)}$. Since
\begin{align*}
\log((1-\delta)^{\mu(1-\delta)+\frac{1}{2}})
&=(\mu(1-\delta)+\frac{1}{2})\log(1-\delta)\\
&=(\mu(1-\delta)+\frac{1}{2})(-\delta-\frac{\delta^2}{2}+o_{\delta\downarrow 0}(\delta^2))\\
&=-\mu\delta+\frac{\mu\delta^2}{2}+o_{\mu\uparrow\infty}(\mu^{\frac{1}{3}}),
\end{align*}
it follows that 
\begin{align}
(1-\delta)^{-\mu(1-\delta)-\frac{1}{2}}=e^{\mu\delta-\frac{\mu\delta^2}{2}+o_{\mu\uparrow\infty}(\mu^{\frac{1}{3}})}.\label{eq:Gapprox}
\end{align}
Substituting (\ref{eq:Gapprox}) into (\ref{eq:stirling}) and taking the limit gives
\begin{align*}
\lim\limits_{\mu\uparrow\infty}(1+a\mu)\lfloor\mu-\mu^{\frac{2}{3}}\rfloor\frac{e^{-\mu}\mu^{\lfloor\mu-\mu^{\frac{2}{3}}\rfloor}}{\lfloor\mu-\mu^{\frac{2}{3}}\rfloor!}=\lim\limits_{\mu\uparrow\infty}(1+a\mu)\mu(1-\delta)\frac{e^{-\frac{\mu\delta^2}{2}+o_{\mu\uparrow\infty}(\mu^{\frac{1}{3}})}}{\sqrt{2\pi\mu}}=0,
\end{align*}
which, together with (\ref{eq:mode}), proves (\ref{eq:term3}).

\end{document}